\documentclass[%
 reprint,
 amsmath,amssymb,
 aps,
prb,
]{revtex4-2}

\usepackage{graphicx}
\usepackage{dcolumn}
\usepackage{bm}

\usepackage[final]{changes}

\usepackage[dvipsnames]{xcolor}
\makeatletter
\AddToHook{cmd/added/before}{\def\Changes@AuthorColor{red}}
\AddToHook{cmd/deleted/before}{\def\Changes@AuthorColor{orange}}
\AddToHook{cmd/replaced/before}{\def\Changes@AuthorColor{blue}}
\makeatother

\usepackage[T1]{fontenc}
\usepackage[utf8]{inputenc}

\begin{document}

\preprint{APS/123-QED}

\title{Electric-dipole-induced nuclear electric resonance in silicon}

\author{Jiyuan Su}
    \altaffiliation{These authors contributed equally to this work.}
    \affiliation{Southern University of Science and Technology, Shenzhen 518055, China}
	\affiliation{International Quantum Academy, Shenzhen 518048, China}
    \author{Shihang Zhang}
    \altaffiliation{These authors contributed equally to this work.}
    \affiliation{Southern University of Science and Technology, Shenzhen 518055, China}
	\affiliation{International Quantum Academy, Shenzhen 518048, China}
	\author{Peihao Huang}  
	\email{huangpeihao@iqasz.cn}
    \affiliation{International Quantum Academy, Shenzhen 518048, China}





\date{\today}

\begin{abstract}
The nuclear spin of a phosphorus atom in silicon is considered as a promising quantum information carrier for quantum computing due to its long coherence time. Coherent control of the nuclear spin is typically realized using nuclear magnetic resonance (NMR). In this work, we theoretically propose that with the aid of hyperfine interactions and magnetic field gradients, fast electric-dipole-induced nuclear electric resonance (EDNER) can be achieved. The stronger EDNER (compared with NMR) arises from the combined effects of hyperfine interaction between the nuclear and electron spins and coupling between the electron spin and charge degrees of freedom mediated by magnetic field gradients in this scheme. By operating at specific working points, the nuclear-spin splitting becomes less sensitive to electrical noise, enabling long coherence times and high-fidelity quantum operations. All elctrical, addressable control of nuclear spins paves the way for scalable silicon-based quantum computing.

\end{abstract}

\maketitle


\section{\label{introduction} Introduction}


The donor-based spin qubit in silicon has attracted wide interest due to its long coherence time, high gate fidelity, and compatibility with modern semiconductor processes~\cite{Kane1998,tyryshkin2012,muhonen_morello_2014storing,Thorvaldson2024,edlbauer2025}.
Experimental results in donor systems have shown the coherence time of electron (nuclear) spin qubits lasting approximately 30 seconds (1 minute) \cite{muhonen_morello_2014storing}. Moreover, single-qubit gates of nuclear spin qubits have exhibited an impressive fidelity of 99.98$\%$, whereas two-qubit gates have achieved a fidelity of 99.9$\%$  \cite{mkadzik_morello_2022precision,Thorvaldson2024,edlbauer2025}. Additionally, a high-speed two-qubit gate of electron spin qubits with a fidelity of 90$\%$  and a duration of 800 ps has also been successfully implemented \cite{he_simmons_2019two}.
These progresses prove that the donor can be used as a carrier of qubits for quantum information processing.

The control of single spins is essential for spin-based quantum computing. 
Single-qubit gate fidelity is typically limited by crosstalk and qubit decoherence. Therefore, addressable and fast electrical control offers a viable route to improving qubit operation performance.
Generally, the electrical control of electron spin qubit is achieved by electric dipole spin resonance (EDSR) based on spin-orbit coupling~\cite{golovach2006electric,Tokura2006}. For silicon, a material with weak intrinsic spin-orbit coupling, EDSR can be enabled by magnetic field gradients~\cite{yoneda_Tarucha_2018quantum,Noiri2022,Xue2022}. This approach substantially enhances the electric dipole strength, enabling Rabi frequency up to 30 MHz and strong spin-photon coupling in gate-defined silicon quantum dots.~\cite{yoneda_Tarucha_2018quantum,Hu2012,Mi2018,borjans2020,Harvey-Collard2022,yu2023,dijkema2025}.

For donor systems, the hyperfine interaction can be electrically modulated to flip the spins of both electrons and nuclei simultaneously with a flip-flop Rabi frequency of 118 kHz \cite{tosi_morello_2017silicon,tosi_morello_2018robust,savytskyy_morello_2022electrically}. This can also be referred to as EDSR of the flip-flop qubit. However, the electrical manipulation of the nuclear spin qubit has not been realized yet. The single nucleus is weakly coupled with the environment and thus has a long coherence time. Nevertheless, the weak coupling also results in slow qubit manipulation and the difficulty in coupling qubits to other quantum systems.
In the case of nuclear magnetic resonance (NMR), the Rabi frequency of spin manipulation is typically limited to tens of kHz \cite{pla_morello_2013high,edlbauer2025}, necessitating fast electrical manipulation of a nuclear spin to improve the fidelity of spin operations.
In addition, the frequency of the nuclear electric quadrupole interaction is approximately 1 kHz \cite{asaad_morello_2020coherent}. A fast and convenient manipulation scheme of individual nuclear spins remains to be illustrated.

Here, we propose a fast electrical manipulation scheme for nuclear spin in a quantum dot (QD)-donor system consisting of an interface QD and a phosphorus (P) donor. In the system, an oscillating electric field can modulate the hyperfine interaction between the electron and nuclear spins, thereby coupling the nuclear spin dynamics to electrically driven electron spin operation. Therefore, all-electrical manipulation of the nuclear spin is achieved by exploiting the EDSR on both electron spin qubit and the flip-flop qubit, a mechanism that we term electric-dipole-induced nuclear electric resonance (EDNER). We verify the feasibility of the proposed mechanism through both analytical and numerical calculations. The Rabi frequency of EDNER can reach up to 39 MHz. In addition, we identify charge-noise-insensitive clock-transition (CT) point at both first and second order, which can suppress the decoherence of spin qubits during electrical manipulation. As a result, high-fidelity operation of nuclear spin qubits can be achieved, providing a pathway toward scalable quantum computing based on donor nuclear spin qubits.













\section{model}

\begin{figure}[t] 
        \centering
        \includegraphics[width=0.8\columnwidth]{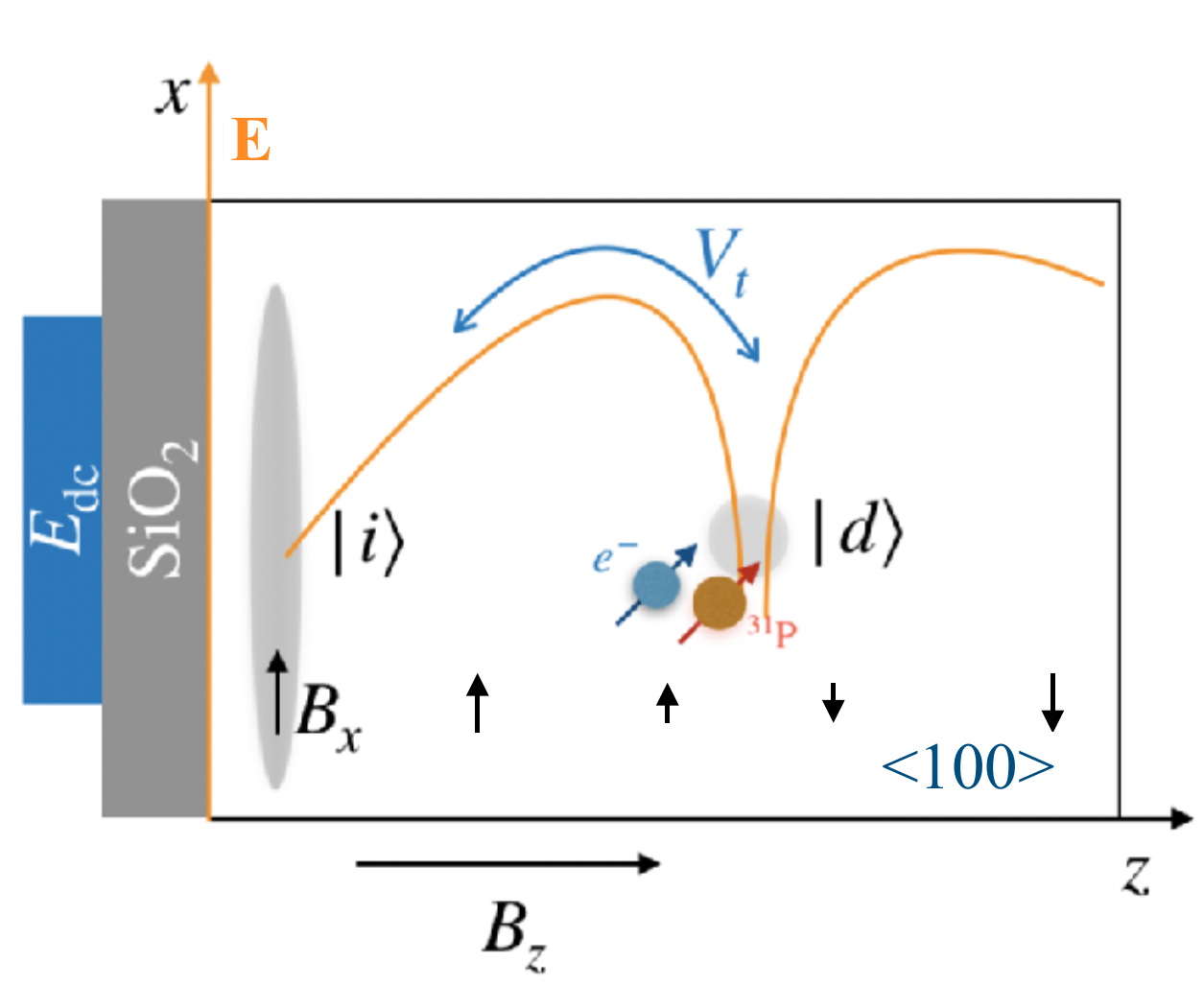}
        \caption{
        \label{fig: device}  Schematic illustration of the quantum dot-donor system with a static magnetic field $B_z$ and an inhomogeneous perpendicular magnetic field $B_x$. The electron interacts with the nucleus through the hyperfine interaction. The vertical electric field $E_{\mathrm{dc}}$, \replaced{applied by the metallic top gates,}{which} controls the electron orbital wavefunction distribution between the bulk-like donor state $|d\rangle$ and interface-like state $|i\rangle $\deleted{, is applied by the metallic top gates}.
        }
\end{figure}

The QD-donor system considered in this work is shown in Fig.~\ref{fig: device}. A $\mathrm{^{31}P}$-donor is implanted into $\mathrm{^{28}Si}$ substrate material, and is coupled to a interface \deleted{gate-defined quantum dot (}QD above~\cite{tosi_morello_2017silicon}.  
The nuclear spin of the P donor serves as a qubit.
There are a static magnetic field $B_z$ and an inhomogeneous perpendicular magnetic field $B_x$.
$B_z$ quantizes the electron and nuclear spin in the $z$ direction via the Zeeman interaction.
Moreover, the inhomogeneous magnetic field $B_x$ couples the electron spin with the orbital degree of freedom providing a mechanism for spin manipulation.
Additionally, an electric field \deleted{$E_{\mathrm{dc}}$ }applied to the top gate of the device can change the \replaced{detuning}{tunneling $V_t$} of the electron between the donor site and the interface QD\deleted{, where the orbital wave function of the electron at the donor site and the interface are denoted as bulk-like donor state $|d\rangle$ and interface-like state $|i\rangle$}.

The system Hamiltonian is given by 
\begin{equation} \label{eq:Htotal1} 
H=H_{\mathrm {orb }} +H_{B_{0}}^{\mathrm { }}+ H_{A}^{\mathrm { }}+H_{B_{1}}^{\mathrm { }}+V_{\mathrm {ext}},
\end{equation} 
where $V_{\mathrm {ext}} $ represents the \added{potential due to interaction with} externally applied driving electric field, which enables control over the electronic orbital wavefunction.
$H_{\mathrm {orb }}$ is the orbital Hamiltonian,
$H_{B_{0}}$ ($H_{B_{1}}$) is the Zeeman Hamiltonian of particles under magnetic field $B_z$ (inhomogeneous magnetic field $B_x$). $H_A$ describes the hyperfine interaction between the electron spin and the nuclear spin. 

The orbital degrees of freedom of an electron are approximated as a two-level system\deleted{ in the orbital Hamiltonian}:
\begin{equation} \label{eq:Horb} 
H_{\mathrm {orb}}= \frac{V_{t}}{2} \tau_{x} -\frac{\epsilon}{2} \tau_{z},
\end{equation} 
where $\tau_{z}=|i\rangle\langle i|-| d\rangle\langle d|$ and $\tau_{x} =|i\rangle\langle d|+| d\rangle\langle i| $ are the Pauli operators for the electron orbital degree of freedom. $|i\rangle$ ($|d\rangle$) represents the charge state of electron at the interface QD (donor). \replaced{$V_{t}$ (and $\epsilon=eE_{\mathrm{dc}} d_0/\hbar$) is tunneling (detuning)}{The detuning $\epsilon = e E_{\mathrm{dc}}d_0/ \hbar $ is the energy} between $|d\rangle$ and $|i\rangle$. $d_0$ is the vertical distance between the donor and the interface quantum dot, $e$ is the electronic charge, and $\hbar$ is the Planck constant. $E_{\mathrm{dc}}$ is the vertical electric field applied by the top metal gate.
At the ionization point ($E_{\mathrm{dc}}$  = $E_{\mathrm{dc}}^{0}$), where the energy detuning is \deleted{close to }zero, the orbital states are strongly hybridized. \added{The ground state is $ |g\rangle=\frac{1}{\sqrt{2}}(|i\rangle-|d\rangle)$ and the first excited state is $|e\rangle=\frac{1}{\sqrt{2}}(|i\rangle+|d\rangle)$}. The  energy difference between the two orbital eigenstates is equal to $V_t$, and \deleted{$\partial \Omega_{0} / \partial \epsilon=0$, where} the charge qubit has a first-order sweet spot \added{with $\partial \Omega_{0} / \partial \epsilon=0$} \cite{kim_microwave-driven_2015} relative to the electric noise. It is beneficial for improving spin coherence as shown later.

$H_{B_0}$ is the Zeeman Hamiltonian describing the interaction with the static magnetic field $B_{z}$ along the $z$-direction:
\begin{equation} \label{eq:Hzeeman} 
H_{B_{0}}= B_{z}\gamma_{e}^{\mathrm {orb}} S_{z}-B_{z}\gamma_{n} I_{z},
\end{equation}
where $I_{z}$ is the $z$-component of the nuclear spin operator, whose eigenstates are $\mid\Uparrow\rangle$ and $\mid\Downarrow\rangle$, and the gyromagnetic ratio is $\gamma_{\mathrm{n}}/2\pi =17.23$ $\mathrm{MHz} / \mathrm{T}$. $S_{z}$ is the $z$-component of the electron spin operator, whose eigenstates are $\mid\uparrow \rangle$ and $\mid\downarrow\rangle$, and the gyromagnetic ratio is $\gamma_{\mathrm{e}}/2 \pi =27.97$ $\mathrm{GHz} /\mathrm{T}$. 
\replaced{Note that the electron gyromagnetic ratio at a Si/SiO$2$ interface may deviate from that of a donor-bound electron by as much as 0.7\%, as quantified by the relative difference $\Delta{\gamma_e}$}{Note, the gyromagnetic ratio of an electron confined at a Si/SiO$_2$ interface can differ from that of a donor-bound electron, with the relative shift by a relative amount $\Delta_{\gamma_e}$ up to 0.7\%} 
\cite{Rahman2009,tosi_morello_2017silicon}. Therefore, $\gamma_e^{\mathrm {orb}}$ is determined by the electron charge state:
\begin{equation} \label{eq:gamme_orb} 
\gamma_e^{\mathrm {orb} }
=\left[1+\left(\frac{1+\sigma_{z}}{2}\right) \Delta_{\gamma_e}\right] \gamma_e.
\end{equation}

$H_{B_1}$ is the Zeeman Hamiltonian for the inhomogeneous magnetic field $B_{x}$, which is equal to zero at the donor position. Under the first-order expansion of the gradient,
\begin{equation} \label{eq:HB1} 
H_{B_{1}}
= \frac{1}{2}b_xd_{0}(1+\tau_z ) \gamma_{e}^{\mathrm {orb}} S_{x},
\end{equation} 
where $b_x$ is the magnetic field gradient. $ \tau_{z}  S_{x}$  hybridizes  the electron spin and orbital state, forming an artificial spin-orbit coupling (a-SOC)\added{, which is important} for EDSR between electron spin states $\mid\uparrow \rangle$  and $\mid\downarrow \rangle$ \cite{yoneda_Tarucha_2018quantum,Tokura2006}.

$H_A$ describes the hyperfine interaction between the electron and the nucleus: 
\begin{equation} \label{eq:H_A} 
H_{A}
=\frac{A_{hf}}{2}\left(1-\tau_{z}\right) \mathbf{S} \cdot \mathbf{I},
\end{equation} 
where $A_{hf}$ is set to 117 $\mathrm{MHz}$~\cite{Feher1959}, $\mathbf{S}$ ($\mathbf{I}$) is the electron (nuclear) spin operator, and $\mathbf{S} \cdot \mathbf{I}=\frac{1}{2}\left(S_{+} I_{-}+S_{-} I_{+}\right)+S_{z} I_{z}$. The term $ \tau_{z}\left(S_{+} I_{-}+S_{-} I_{+}\right) $ couples the the electron spin and nuclear spin to the orbital degree of freedom to allow EDSR to be achieved between the flip-flop qubit states  $\mid\uparrow \Downarrow\rangle$  and $\mid\downarrow \Uparrow\rangle$. When the electron is fully bound to the nucleus ($|d\rangle$ state), the hyperfine interaction reaches the maximum value ($A_{hf}$). When the electron is tunneled to the interface by adjusting $E_{\mathrm{dc}}$, the hyperfine interaction is weakened.

In our proposal, hyperfine interaction is identified as the driving mechanism for flip-flop qubits, and magnetic field gradient is for the EDSR driving of electron spin qubits. Combining the two driving mechanisms yields a \deleted{second-order Raman-type} process for EDNER as discussed in the following section.

\section{Result}

To demonstrate the principle and performance of EDNER based on a Raman-type process, we perform numerical calculation on the Hamiltonian to analyze its eigenstates. Based on the eigenstates, we obtained the transition frequency of the nuclear spin and the strength of the electric dipole and further evaluated the Rabi frequency and dephasing rate of the \added{nuclear spin} qubit. The numerical results demonstrate that the hybridization of nuclear spin, electron spin, and charge in this scheme leads to a electric dipole transitions up to 40 MHz. Although hybridization also makes the system highly sensitive to electrical noise, the nuclear spin qubit acquires the maximum electric dipole driving strength while \replaced{its resonance frequency remains insensitive to electrical noise to first order}{being less sensitive to electrical noise in its resonance frequency}, near the ionization point. The qubit can achieve over 100 Rabi oscillations before qubit dephasing, enabling high-fidelity quantum gates.

\subsection{\replaced{Qubit definition}{EDNER}}

\begin{figure}[!t] 
        \centering
        \includegraphics[width=0.7\columnwidth]{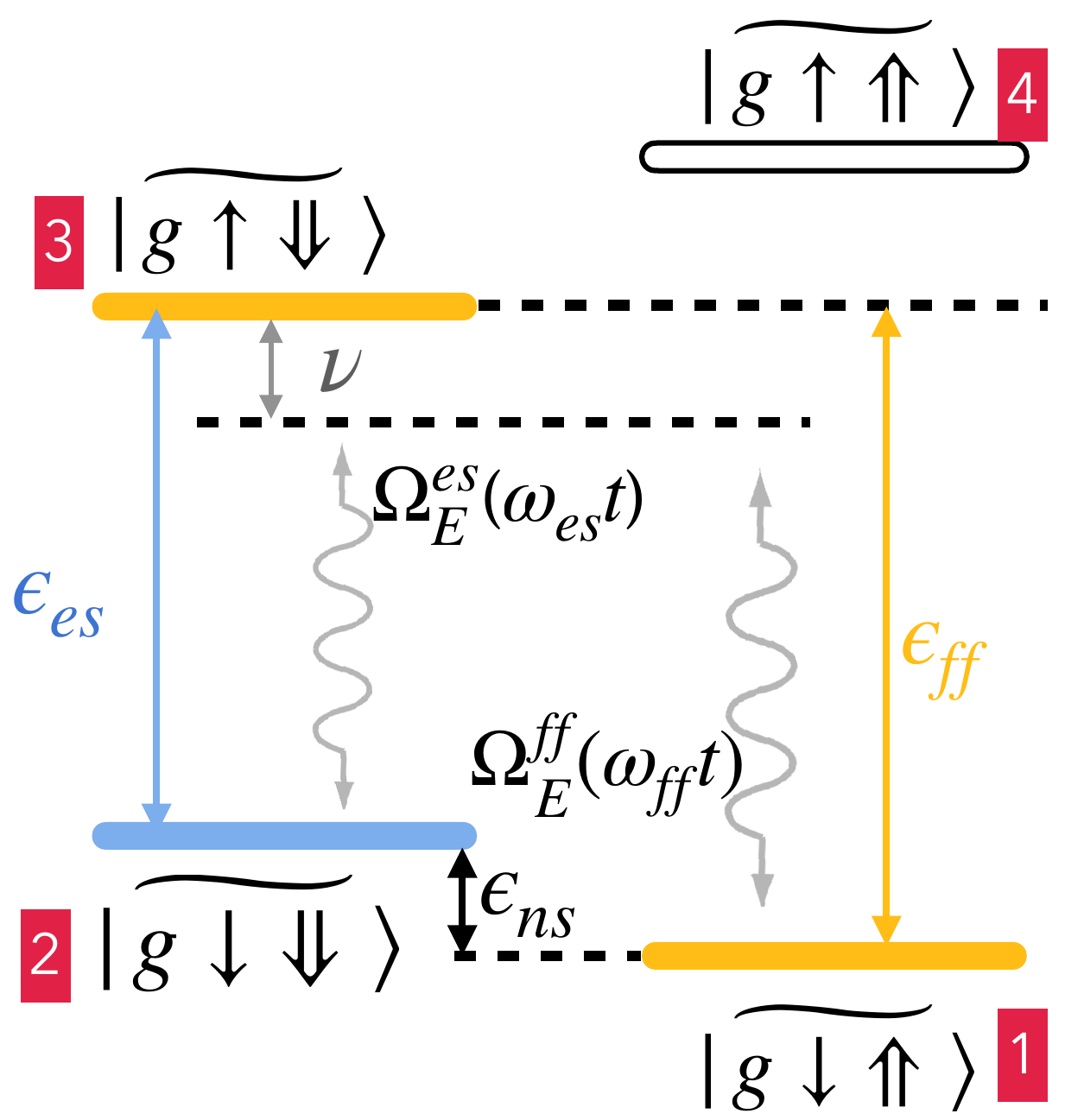}
        \caption{
        \label{fig: Ramanlevel}  Energy level diagram for Raman-enable EDNER of the Si:P nuclear spin qubit. $\epsilon_{es}$ and $\epsilon_{ff}$  are the driving frequency of the transitions $|g  \uparrow \Downarrow \rangle   \leftrightarrow |g  \downarrow \Downarrow \rangle$  and $|g  \uparrow \Downarrow \rangle   \leftrightarrow |g  \downarrow \Uparrow \rangle $, while $\Omega_{E}^{es}$  and $\Omega_{E}^{ff}$  are the respective coupling strengths. The \replaced{two driving frequencies}{two couplings} are equally detuned $\nu$ from their respective transitions. EDNER with energy splitting $\epsilon_{ns}$ is achieved by inducing the coupling of two low-energy states  $|g  \downarrow \Downarrow \rangle$ and $|g  \downarrow \Uparrow \rangle $ through the higher state $|g  \uparrow \Downarrow \rangle $ by combining electron-spin \added{EDSR} and flip-flop EDSR. We \replaced{chose}{have} a fixed value of $\omega_{ff}=20$ GHz, and $\omega_{es}$ is chosen as $\omega_{ff}-\epsilon_{ns}(E_{dc})$.
        }
\end{figure}


The system comprises three degrees of freedom: electron spin, nuclear spin, and charge orbit, resulting in an eight-dimensional Hilbert space. 
In the absence of an external drive, the two lowest eigenstates $\widetilde{|g  \downarrow \Uparrow \rangle}$ and $\widetilde{|g  \downarrow \Downarrow \rangle}$ are defined as the basis of the qubit.
As shown in Fig.~\ref{fig: Ramanlevel}, there is no direct electric coupling between $\widetilde{|g  \downarrow \Uparrow \rangle}$ and $\widetilde{|g  \downarrow \Downarrow \rangle}$. Nevertheless, they can couple to the virtual energy level (in a rotating frame) detuned by $\nu$ from $\widetilde{|g \uparrow \Downarrow\rangle}$.
$\epsilon_{es}$ and $\epsilon_{ff} $ are the resonant frequencies of the transitions $\widetilde{|g  \uparrow \Downarrow \rangle}  \leftrightarrow \widetilde{|g  \downarrow \Downarrow \rangle}$  and $\widetilde{|g  \uparrow \Downarrow \rangle }  \leftrightarrow \widetilde{|g  \downarrow \Uparrow \rangle }$, while $\Omega_{E}^{es}$ and $\Omega_{E}^{ff}$  are the corresponding coupling strengths, and $\omega_{es}$ and $\omega_{ff} $ are the corresponding driving frequencies for EDNER.
Nuclear electric resonance  with energy splitting $\epsilon_{ns}$   is achieved by inducing the coupling of states $\widetilde{|g  \downarrow \Uparrow \rangle}$ and $\widetilde{|g  \downarrow \Downarrow \rangle}$ through the virtual level. The nuclear dipole transition strength is
\begin{equation}
    \Omega^{ns}_E=-\frac{1}{ v} \Omega^{es}_E \Omega^{ff}_E.
\end{equation}
The condition for this indirect coupling  is that the coupling strength  $\Omega_{E}^{es}$ and $\Omega_{E}^{ff}$ are \replaced{much less than}{small compared to} the detuning $v$, and  $\omega_{es}=\omega_{ff}- \epsilon_{ns}(E_{dc})$. 

\begin{figure}[!t] 
        \centering
        \includegraphics[width=0.7\columnwidth]{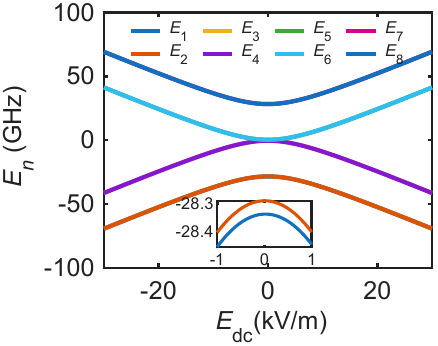}
        \caption{
        \label{fig: energy level}  Energy levels  $E_n$ ($n$ = 1,...,8) as a function of the static vertical electric field  for  $V_t/h=28.7$ $\mathrm{GHz}$, $B_z=1$ $\mathrm{T}$,  $b_x=0.5$ $\mathrm{mT/nm}$, $\Delta_{\gamma_e }=-0.3\%$, $d_{0}=14.72$ nm. The inset shows the lowest two eigenvalues to distinguish the nuclear spin states.
        }
\end{figure}

To understand the mechanism and performance of the \replaced{EDNER}{nuclear spin transition}, we numerically diagonalize the Hamiltonian of the system in an eight-dimensional Hilbert space with charge, electron spin, and nuclear spin ground states $\{|e\rangle,|g\rangle\} \otimes \{\mid\uparrow\rangle,\mid\downarrow\rangle\} \otimes \{\mid\Uparrow\rangle,\mid\Downarrow\rangle\}$. Diagonalizing the time-independent Hamiltonian yields eight eigenstates $|\psi_n\rangle $ and eight eigenvalues $E_n$ ($n = 1,\ldots,8$), ordered \replaced{in}{by} increasing energy, as shown in  Fig.~\ref{fig: energy level}. 
Analyzing the composition of the bare states in the eigenstates, we found that the lowest energy level $| \psi_2 \rangle$ ($| \psi_1 \rangle$) is mainly occupied by the bare state $| g  \downarrow \Downarrow \rangle$ ($| g \downarrow \Uparrow \rangle$),
since the coupling between the two lowest energy eigenstates and the higher energy orbitals is weak. \added{In this study,} the two lowest-energy states define the qubit subspace.


The introduction of an effective spin-orbit coupling makes the qubit frequency sensitive to the electric field. Figure 4 shows the dependence of the qubit frequency on the electric field $E_{\mathrm{dc}}$. When $E_{\mathrm{dc}}$ is small ($-30$ kV/m), the electron is fully bound to the donor nucleus, and the corresponding hyperfine interaction reaches its maximum value (117 MHz). As the electric field increases, the electron wavefunction gradually shifts toward the interface QD, leading to a reduction in the hyperfine interaction strength, and thereby a decrease in the qubit frequency. \replaced{When}{In the region where} the electric field is close to \replaced{$E_{dc}^{0}$}{zero}, the \added{energy of} electron charge states is most sensitive to the electric field, resulting in the most pronounced variation of the qubit frequency. When the electron becomes localized in the interface QD, the qubit frequency saturates toward its minimum value. The dependence of the qubit frequency on the electric field reflects its sensitivity to charge noise and can significantly affect qubit dephasing, as discussed in \replaced{Sec.~\ref{sec:performance}}{the detail later}.


\begin{figure}[!t] 
        \centering
        \includegraphics[width=0.7\columnwidth]{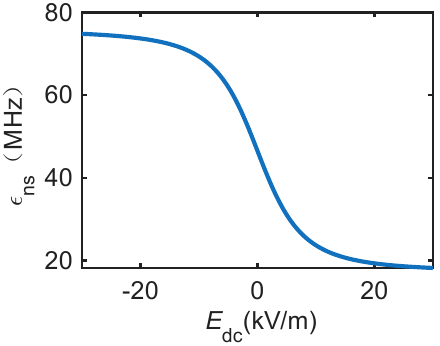}
        \caption{
        \label{fig: nuclear energy }  Nuclear spin qubit transition frequency. As the $E_{\mathrm{dc}}$ approaches \replaced{$E_{dc}^{0}$ ($E_{dc}^{0}=0$ is defined for simplicity)}{0}, the nuclear spin frequency becomes highly sensitive to the $E_{\mathrm{dc}}$.
        }
\end{figure}


\subsection{EDSR for Flip-flop qubit and electron spin qubit}

As shown in Fig.~\ref{fig: Ramanlevel}, EDNER is achieved via the intermediate state $\widetilde{|g  \uparrow \Downarrow \rangle}$ by combining flip-flop EDSR and electron spin EDSR. For simplicity, we denote the flip-flop EDSR between $|\psi_1\rangle$ and $\widetilde{|g  \uparrow \Downarrow \rangle}$ as \replaced{ff-EDSR}{EDSR$_1$}, and the electron-spin EDSR between $|\psi_1\rangle$ and $\widetilde{|g  \uparrow \Downarrow \rangle}$ as \replaced{es-EDSR}{EDSR$_{2}$}. In this section, we analyze the role of the intermediate state $\widetilde{|g  \uparrow \Downarrow \rangle}$ and evaluate the performance of these \replaced{types of}{two} EDSR mechanisms in the system.

\begin{figure}[!t] 
        \centering
        \includegraphics[width=1\columnwidth]{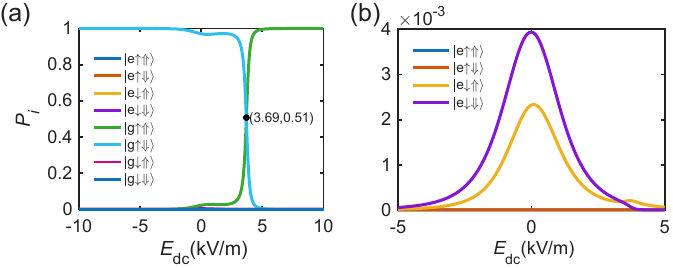}
        \caption{
        \label{fig: Eigenstate3} Proportions $P_{i}$ of the basis states in $|\psi_3\rangle$. (a) Proportions of all eight basis states. Near the crossing-over point, the proportion of bare states changes because the hyperfine interaction is also a function of $E_{\mathrm{dc}}$. The crossing of bare states will lead to high Rabi noise. (b) Proportions of the four charge-excited states.
        }
\end{figure}

Since the degree of hybridization directly affects the EDSR strength, we first calculate the proportions of the basis states in $|\psi_3\rangle$, as shown in Fig.~\ref{fig: Eigenstate3}.
In the region around $E_{\mathrm{dc}}=0$, the eigenstate is represented as a superposition of orbital ground and excited states, resulting from the strong coupling of spin and orbit when the orbital detuning is zero. In this region, $| \psi_3 \rangle$ is mainly occupied by $|g\uparrow \Downarrow\rangle$ (blue line), but with some fraction of state $|g \uparrow \Uparrow\rangle$ (green line), $|e \downarrow \Downarrow\rangle$ (purple line) and $|e \downarrow \Uparrow\rangle$ (yellow line). 
\deleted{The transition between states $|\psi_1 \rangle$ and $| \psi_3 \rangle$ is driven by EDSR$_1$. Therefore,} The dipole strength \added{of ff-EDSR between $|\psi_1 \rangle$ and $| \psi_3 \rangle$} depends on the proportion of the  bare states $|g  \downarrow \Uparrow \rangle$ in $| \psi_1 \rangle$ and   $|e \downarrow \Uparrow \rangle$ in $| \psi_3 \rangle$. 
For \replaced{es-EDSR}{EDSR$_{2}$} between states $|  \psi_2 \rangle$ and $|  \psi_3 \rangle$, the dipole strength depends on the proportion of the  bare states $|g  \downarrow \Downarrow \rangle$ in $|  \psi_2 \rangle$ and   $|e \downarrow \Downarrow \rangle$ in $| \psi_3 \rangle$. 

\begin{figure}[!h] 
        \centering
        \includegraphics[width=1\columnwidth]{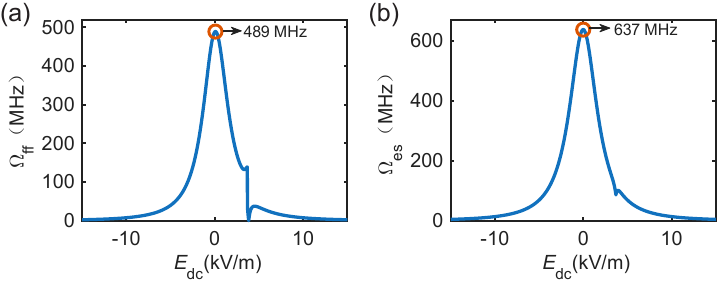}
        \caption{
        \label{fig: Edc<3690 dipole_es&dipole_ff}  The electric dipole strength of flip-flop qubit and electron spin qubit. The strong spin-orbit coupling near $E_{\mathrm{dc}}=0$ results in a strong electric dipole moment. (a) The electric dipole strength $\Omega_{\mathrm{ff}}$ of the flip-flop qubit is plotted as a function of the electric field $E_{\mathrm{dc}}$, with a maximum value of 489 MHz. A dip appears near the CP as the dominant bare-state components of $| \psi_3 \rangle$ and $| \psi_4 \rangle$ are exchanged. (b) The electric dipole strength $\Omega_{\mathrm{es}}$ of the electron spin qubit is plotted as a function of $E_{\mathrm{dc}}$, with a maximum value of 637 MHz.}
\end{figure}

\replaced{Fig.}{Figure}~\ref{fig: Eigenstate3} shows that the dominant component of states $| \psi_3 \rangle$  changes near the crossing-over point (CP), which is at $E_{\mathrm{dc}}=3.69$ kV/m. 
The bare-state energies depend on the \added{effective} hyperfine interaction \added{between electron and nuclear spins}, whose strength is determined by the electron orbital state. As the electric field $E_{\mathrm{dc}}$ increases, the electron is gradually \replaced{moved}{displaced} way from the donor nucleus, which reduces the \added{effective} hyperfine interaction. 
\replaced{Consequently, the relevant energy levels transition from the hyperfine-dominated regime to the Zeeman-dominated regime. This transition rearranges the bare-state composition of the eigenstates.}{Consequently, the relevant energy levels evolve from being dominated by hyperfine interaction to being mainly determined by the Zeeman term, leading to a rearrangement of the dominant bare-state proportions in the eigenstates.} 
Near the CP, the charge state is strongly tunable by $E_{\rm{dc}}$, making the qubit highly sensitive to charge noise and leading to \replaced{increased}{enhanced} qubit decoherence. Depending on the electric field, different states are involved in the calculation of the EDSR coupling: 
\begin{equation}\label{FF dipole}
\Omega^{ff}_E=\left\{\begin{array}{c}\left\langle  \psi_1  \left|  V_{\mathrm{ext}}\right|    \psi_3  \right\rangle , \quad  E_{\mathrm{dc}}<\text{CP} \\ \left\langle  \psi_1  \left|  V_{\mathrm{ext}}\right|    \psi_4  \right\rangle, \quad E_{\mathrm{dc}}\geq \text{CP} \end{array}\right.
\end{equation}
\begin{equation}\label{ES dipole}
\Omega^{es}_E=\left\{\begin{array}{c}\left\langle  \psi_3  \left|  V_{\mathrm{ext}}\right|    \psi_2  \right\rangle , \quad E_{\mathrm{dc}}< \text{CP} \\ \left\langle  \psi_4  \left|  V_{\mathrm{ext}}\right|    \psi_2  \right\rangle, \quad E_{\mathrm{dc}}\geq \text{CP} \end{array}\right.
\end{equation}
where $V_{\mathrm{ext}}$ represents the interaction with the ac electric field $E_{ac}$. \replaced{Hyperfine-mediated flip-flop transitions, together with magnetic-field-gradient-induced EDSR, couple the spin states to the charge degree of freedom, thereby enabling electrical control of the hybridized electron-spin and flip-flop states.}{Hyperfine interactions and magnetic field gradients couple the spin and charge degrees of freedom, thus allowing the use of electric fields to control spin states.}
\begin{equation} \label{Vext} 
    V_{\mathrm{ext}} \approx-\vec{r} \cdot \vec{F}_{\mathrm{ext}} (t) \approx F_{\mathrm{ext},z} (t)d\sigma_{z}^{} = E_{\mathrm{ac}}(t)\sigma_{z}.
\end{equation}
where $E_{\mathrm{ac}}=10\ \mathrm{GHz}$ ($F_{\mathrm{ext},z}=2.8$ kV/m) is the strength of the applied ac driving electric field. The calculated EDSR dipole strength is shown in Fig.~\ref{fig: Edc<3690 dipole_es&dipole_ff}.

Under the ac electric field $E_{\mathrm{ac}}$, the dipole strength of the flip-flop qubit can reach 489 MHz and that of the electron spin can reach 637 MHz around the ionization point.
Around the CP, the changes in the proportion of the bare states within the eigenstates lead to fluctuations in $\Omega^{ff}_E$ and $\Omega^{es}_E$, resulting in steep\deleted{er} curves and enhanced sensitivity to charge noise, thereby causing strong decoherence as discussed in Sec.~\ref{sec:performance}. At the same time, the \replaced{ff-EDSR}{EDSR$_{1}$} coupling strength is substantially suppressed.

\subsection{EDNER \added{performance} for nuclear spin}\label{sec:performance}

\begin{figure}[!t] 
        \centering
        \includegraphics[width=0.7\columnwidth]{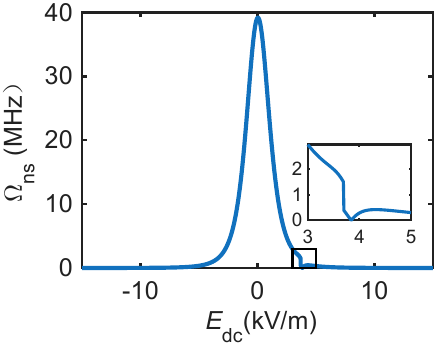}
        \caption{
        \label{fig:dipole_ns} Electric dipole strength of the nuclear spin. In the calculation, we assume $\nu = \epsilon_{ff}(E_{\mathrm{dc}})- \omega_{ff}> 16 \Omega_{ff}>12\Omega_{es}$. At $E_{\mathrm{dc}}=$ 40 V/m, the electric dipole strength of the nuclear spin resonance reaches a maximum of 39.19 MHz. Around 3.84 kV/m, the distortion of the waveform caused by the crossing  of the bare-state components in the eigenstate is also evident. Inset: Enlarged view of $\Omega_{\mathrm{ns}}$ near 3.84 kV/m, corresponding to the region marked by the black box in the main panel.}
\end{figure}

Based on the analysis of the two EDSR mechanisms above, we now analyze the EDNER of the qubit. \added{Fig.~\ref{fig:dipole_ns} shows} the EDNER coupling strength $\Omega_{ns}$\deleted{ is shown in Fig.~\ref{fig:dipole_ns}}. At the ionization point, $\Omega^{ns}_E$ reaches a maximum of 40 MHz. The strength of EDNER is three orders of magnitude stronger than that of NMR (typically $\sim$ 30 kHz) \cite{pla_morello_2013high} and four orders of magnitude faster than the strength of the Rabi frequency achieved using nuclear electric quadrupole interactions (approximately 1 kHz) \cite{asaad_morello_2020coherent}. 
As discussed above, near the CP, the change in the dominant bare-state proportions within the eigenstates causes the strength of \replaced{ff-EDSR}{EDSR$_1$} to approach zero, which suppress the corresponding EDNER strength. \replaced{We next evaluate the qubit dephasing rates to identiy the optimal operating point.}{For optimal control, the system should therefore be operated away from the CP.}

Electrical control based on the mechanism like EDSR might introduce additional dephasing mechanisms, thereby enhancing qubit decoherence. To assess the impact of EDNER on qubit decoherence, we calculate the charge-noise-induced qubit dephasing rate. Two main mechanisms contribute to qubit dephasing. one arises from charge-noise-induced fluctuations of the qubit frequency mediated by the aSOC and the hyperfine interaction. For the distance between the donor and the interface QD $d_{0} \sim$14nm, the potential fluctuation would correspond to an r.m.s. \added{vertical electric-field fluctuation}. Assuming a charge-noise-induced electric field fluctuation $\delta E$ of 100 V/m, the  nuclear spin dephasing rate is 
\begin{equation}\label{1/T2}
    \frac{1}{T^{*}_{2}}= \frac{\partial \epsilon_{ns}} {\partial E_{\mathrm{dc}} } \delta E_{z}.
\end{equation}
The other \added{dephasing mechanism} originates from charge-noise-induced fluctuations of the EDSR coupling strength, leading to dephasing of the driven qubit:
\begin{equation}\label{1/T2*}
    \frac{1}{T_{2,\Omega}}= \frac{\partial D_{ns}} {\partial E_{\mathrm{dc}} } \delta E_{z}.
\end{equation}

\begin{figure}[!t] 
        \centering
        \includegraphics[width=1\columnwidth]{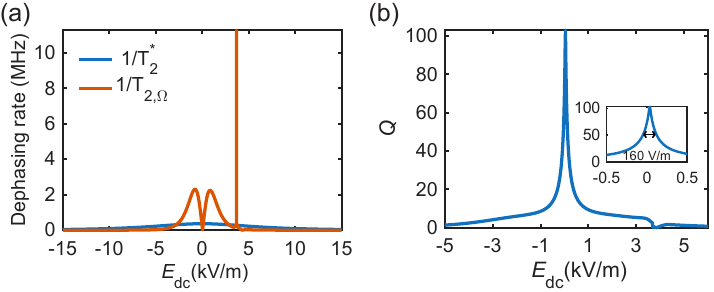}
        \caption{
        \label{fig: dephasing and Q}  (a) Comparison of dephasing rates induced by fluctuations in the qubit frequency (blue solid line) and EDNER driving strength (red solid line).  (b) Quality factor $Q$ \replaced{in presence of dephasing}{under both dephasing mechanisms}. Inset: quality factor $Q$ over the range $-0.5$ and $0.5$ kV/m. The full width at half maximum is 160 V/m.}
\end{figure}

\replaced{Fig.}{Figure}~\ref{fig: dephasing and Q}(a) compares the \replaced{two dephasing rate $1/T_2^*$ and $1/T_{2,\Omega}$}{rates of the two dephasing mechanisms}. Nuclear spin dephasing rate $1/T_{2}^{*}$ (blue line) ranges from 1 kHz to 1 MHz.
The maximum \added{dephasing rate} occurs at $E_\mathrm{dc}=$ 100 V/m due to the high sensitivity of the hyperfine interaction to the electric field near the ionization point, leading to charge-noise-induced \replaced{nuclear spin}{qubit} dephasing. The dephasing rate of the driven qubit $1/T_{2,\Omega}$ (red line) has several sweet spots (SSs). 
The first SS occurs at $E_\mathrm{dc}\sim$0, corresponding to the charge SS of the orbital degree of freedom. At this point, the nuclear-spin electric dipole strength $\Omega_{\mathrm{ns}}$ reaches its maximum. The second sweet spot occurs at $E_\mathrm{dc}=$3.84 kV/m (CP) due to a change in the bare state that predominantly occupies $| \psi_3 \rangle$. However, the electric dipole strength of the nuclear spin is substantially suppressed \added{at the second SS}.

One of the key indicators of a  quantum system is the precision of quantum gates, represented by the quality factor $Q$. In this work, the quality factor for X gate is represented by
\begin{equation}\label{Eq: Q}
Q=\Omega_{ns} T_2, 
\end{equation}
where $T_2$ is taken as the smaller of $T_{2}^{*}$ and $T_{2,\Omega}$. As shown in Fig.~\ref{fig: dephasing and Q}(b), $Q$ reaches the maximum \replaced{of about 100 with}{at} $E_\mathrm{dc}\sim$0, \deleted{with a value of about 100 }due to the SS and the strong electric dipole coupling. Moreover, the full width at half maximum of $Q$ is about 160 V/m, which exceeds the strength of the quasi-static electrical noise. \added{Within the typical strength of electrical noise, high-fidelity control of nuclear spin can be achieved at the SS.}


\section{Discussion}
Recently, nuclear spin qubit devices based on donor clusters have achieved remarkable experimental progress, including high-fidelity two-qubit gates~\cite{edlbauer2025}, the preparation of eight-qubit entangled states~\cite{edlbauer2025}, and demonstrations of quantum error detection\deleted{ and correction}~\cite{ZhangCH2025}. A potential bottleneck for the scalability of this system is crosstalk error arising from \added{frequency} crowding. The manipulation scheme proposed in this work can significantly suppress such crosstalk through localized control of both nuclear spin and electron spin qubits. In addition, the fast manipulation of nuclear spins further ensures high-fidelity single-qubit gates.

In general, all-electrical control schemes involve inherent trade-offs. On the one hand, they enable fast and addressable qubit operation. Additionally, the strong electric dipole can induce long-range electric dipole-dipole interactions for multi-qubit gates, thereby reducing the fabrication complexity. On the other hand, electrical control of spin inevitably introduces spin-orbit coupling, making qubit coherence more susceptible to charge noise. As introduced in this work, charge noise can lead to qubit decoherence through its effects on both the qubit frequency and the driving strengths. By adjusting the electric field $E_{\mathrm{dc}}$ to the ionization point, the dephasing rate $1/T_{2,\Omega}$ corresponding to the driving strength can be suppressed while maintaining high-speed operation, thereby enabling high-fidelity qubit gates. However, in the absence of additional spin-orbit coupling, the decoherence time of the $^{31}$P nuclear spin can exceeds 1 ms. This indicates significant potential for further improving qubit coherence in the EDNER scheme, warranting additional investigation \replaced{on qubit dephasing suppression}{in future work}.

Building on the EDNER scheme proposed in this work, an all-electrical control scheme for nuclear spin qubits may be developed\deleted{ in the future}. Several key open directions remain, including the implementation of two-qubit gates via long-range electric dipole-dipole interactions, the extension from single-donor systems to multi-donor clusters, and the large-scale scalability of qubit devices based on the donor-\deleted{interface-}QD structures.

\section{Conclusion}
In \replaced{conclusion}{this paper}, we propose an all-electrical manipulation scheme for nuclear spin based on EDNER. At a driving electric field of 2.8 kV/m and a doping depth of 14.7 nm for phosphorus atoms, the electric dipole strength of nuclear spin can reach 39 MHz, which \replaced{is}{are} three orders of magnitude larger than the NMR strength and four orders of magnitude larger than that of the \replaced{NER}{EDSR} achieved by the electric quadrupole interaction of the nucleus. When the vertical electric field is adjusted to the ionization point, the qubits can achieve more than 100 Rabi oscillations \replaced{within}{during} coherence time. The high-speed, high-fidelity EDNER scheme offers a promising route toward scalable silicon-based quantum computing.




\bibliography{apssamp}

\end{document}